\documentclass[12pt,a4paper]{iopart}
\usepackage{amssymb,graphicx,cite}
\usepackage[usenames]{color}
\eqnobysec

\newcommand{\beq}{\begin{equation}}
\newcommand{\eeq}{\end{equation}}
\newcommand{\beqa}{\begin{eqnarray}}
\newcommand{\eeqa}{\end{eqnarray}}
\newcommand{\ba}{\begin{array}}
\newcommand{\ea}{\end{array}}

\begin{document}

\title[Effective Nonlinear Schr\"odinger Equations at Unitarity]
{Effective Nonlinear Schr\"odinger Equations for Cigar-Shaped 
and Disk-Shaped Fermi Superfluids at Unitarity}

\author{
S. K. Adhikari$^{1}$\footnote{adhikari@ift.unesp.br;
URL: www.ift.unesp.br/users/adhikari}
and L. Salasnich$^{2}$\footnote{salasnich@pd.infn.it;
URL: www.padova.infm.it/salasnich}}
\address{
$^1$Instituto de F\'{\i}sica Te\'orica, UNESP - S\~ao Paulo State
University, 01.405-900 S\~ao Paulo, S\~ao Paulo, Brazil \\
$^2$CNR-INFM and CNISM, Research Unit of Padova,
Department of Physics ``Galileo Galilei'', University of Padua, 
Via Marzolo 8, 35131 Padova, Italy}

\begin{abstract}
In the case of tight transverse confinement (cigar-shaped trap)
the three-dimensional (3D) nonlinear Schr\"odinger equation,  
describing superfluid Fermi atoms at unitarity 
(infinite scattering length $|a|\to \infty$), is reduced to 
an effective one-dimensional form by averaging over the 
transverse coordinates. The resultant 
effective equation is a 1D nonpolynomial Schrodinger equation, 
which produces results in good agreement with the original 3D one. 
In the limit of small and large 
fermion number $N$ the nonlinearity is of simple power-law type. 
A similar reduction of the 3D theory to a two-dimensional form is 
also performed for a tight axial confinement 
(disk-shaped trap). The resultant effective 2D nonpolynomial equation 
also produces results in agreement with the original 3D 
equation and has simple power-law nonlinearity for small and large $N$. 
For both cigar- and disk-shaped  superfluids 
our nonpolynomial Schr\"odinger equations are quite attractive for 
phenomenological application.  
\end{abstract}

\pacs{05.30.Fk,71.10.Ay, 03.75.Ss,67.85.Lm}

\maketitle

\section{Introduction} 
\label{I}

In the last few years several experimental groups have observed 
the crossover \cite{CROV} from the weakly paired 
Bardeen-Cooper-Schrieffer 
(BCS) state to the Bose-Einstein condensation (BEC) 
of molecular dimers with ultra-cold two-hyperfine-component Fermi
vapors of $^{40}$K atoms \cite{greiner,regal,kinast} 
and $^6$Li atoms \cite{zwierlein,chin}. 
The unitarity limit of the Fermi superfluid was attained 
by manipulating an external background magnetic field 
near a Feshbach resonance which allows an experimental 
realization of infinitely large value of the s-wave 
scattering length $a$ \cite{stringa-fermi}. 

{The most interesting feature of the BCS-BEC crossover, 
noted in
experiments \cite{uexp} on a BCS superfluid as well as demonstrated in 
theoretical model calculations \cite{uth} 
is that, due to a dominance of the Pauli repulsion among fermionic 
atoms  over the interatomic attraction, 
the Fermi superfluid remains essentially repulsive 
in the unitarity limit, i.e. the gas does not collapse 
and its properties are quite regular.} 
Elaborated Monte Carlo calculations  have 
confirmed this effect \cite{mc}. A similar conclusion follows from 
an examination of the compressibility of a Fermi gas \cite{baker}. The 
Fermi system should exhibit universal behavior in the unitarity limit which 
should limit the maximum attractive force to a finite value as $a\to 
\pm \infty$  \cite{uexp}. This phenomenon has greatly enhanced 
the interest in the theoretical study of 
a Fermi gas in the unitarity limit \cite{fguni}. 
 
Close to the critical temperature a Fermi superfluid can be
studied by using the Ginzburg-Landau theory with a complex 
order parameter \cite{ginzburg,landau2,leggett}. 
Recently a nonlinear Schr\"odinger (NLS) equation for the complex 
order parameter has been proposed at zero-temperature 
to study the BCS-BEC crossover 
\cite{kim-zubarev,manini05,sala-josephson,ska1}. This equation 
has a simple nonlinear term in the weak-coupling BCS limit 
($a$ negative and small), 
at unitarity ($a = \pm \infty$), and the BEC regime ($a$ positive and 
small) where it becomes the Gross-Pitaevskii equation for 
Bose-condensed molecules.  However, in the full BCS-BEC crossover 
the 3D NLS equation has a complicated nonlinear term 
\cite{stringa-fermi,mc,manini05,HY}. In addition, this 3D NLS  
equation must be solved numerically 
\cite{kim-zubarev,manini05,sala-josephson}. 
Hence an effective one-dimensional (1D) and two-dimensional (2D) 
reduction of this equation under convenient trapping conditions of 
cigar and disk-shaped traps is well appreciated.  The present paper
addresses this important question of 
1D and 2D reduction of the original 3D NLS equation at unitarity. 
Through numerical studies we also investigate the validity of
the 3D-1D and 3D-2D reduction of the 3D NLS equation. 
 
In Sec. \ref{II} we consider a confined 3D Fermi superfluid at unitarity 
and discuss the 3D NLS equation \cite{sala-josephson,sala-new}. 
We derive it from an energy functional which takes into account 
the bulk properties of the system and also the 
inhomogeneities in the density profile due to the external potential. 
We show that the gradient term of our energy functional 
is practically the same of the one recently deduced \cite{rupak} 
with an epsilon expansion of energy density in $4-\epsilon$ 
dimensions \cite{rupak,son}. 
{In addition, we find that 
the total energy calculated from this 
equation for a 3D system of spin 1/2 fermions, 
trapped in a spherically-symmetric harmonic potential,  
is in good agreement \cite{sadhan}
with those obtained from accurate Monte Carlo 
calculations \cite{FN,FN1}.} In Sec. \ref{III} we introduce a generic 
axially-symmetric harmonic potential which models the confining trap 
of the superfluid fermionic system. 
In Sec. \ref{IV} we consider a cigar-shaped 
Fermi superfluid. Using a variational ansatz we minimize the 3D energy 
functional and derive an effective 1D nonpolynomial 
Schr\"odinger (1D NPS) equation for this system. Simple analytic forms 
of the 1D NPS model are derived for small 
and large number of atoms. The variation of the nonlinear term from 
small to large number of atoms is illustrated.  In Sec. \ref{V} we 
consider a disk-shaped 
Fermi superfluid. Using a variational ansatz we minimize the 3D energy
functional and derive an effective 2D nonpolynomial Schr\"odinger  
(2D NPS) equation for this system. 
Simple analytic forms of the 2D NPS model are derived for small 
and large number of atoms. 
In Sec. \ref{VI} we present a numerical study of the different 
model equations in one and two dimensions for cigar- and disk-shaped 
systems and compare the results with the full 3D NLS equation. 
The results of the approximate models are found to be in good 
agreement with exact calculations. 
Finally, in Sec. \ref{VII} we present concluding remarks.  

\section{Fermi superfluid at unitarity}
\label{II}

We consider a dilute Fermi gas of $N$ atoms with two equally
populated spin components and attractive inter-atomic strength.
At zero temperature the gas is fully superfluid and 
the superfluid density coincides with the total 
density. A description of the superfluid
state can be obtained by using a complex order parameter 
$\Psi({\bf r})$ conveniently normalized to the total number 
of superfluid pairs $N_p$ \cite{ginzburg,landau2,leggett}, i.e.
\beq 
\int |\Psi({\bf r})|^2 \ d^3{\bf r} = N_p = {N\over 2} \; .
\label{norma}
\eeq
The local number density of superfluid pairs is defined 
as  $n_p({\bf r}) = |\Psi({\bf r})|^2 = n({\bf r})/2 $ 
where $n({\bf r})$ is the total density (the density of atoms). 
{Notice that $\Psi$ is not the condensate 
wave function, because the modulus
square of it gives the number of particles, not of condensate
particles \cite{sala-josephson,sala-new}.  
At unitarity ($a=\pm \infty$) one finds the condensate density 
of pairs is $n_0 \simeq 0.7 \ n_p$, where $n_p$ is the total 
density of pairs \cite{sala-odlro,giorgini-odlro}. In the BEC side ($a>0$) 
of the BCS-BEC crossover $n_0$ reaches $n_p$ quite 
rapidly by reducing the positive scattering length $a$, 
while in the BCS side ($a<0$) of the crossover $n_0$ 
decreases exponentially to zero by reducing the absolute value 
of the negative scattering length \cite{sala-odlro,giorgini-odlro}. 
However, the phase of $\Psi({\bf r})$ is exactly the phase of the condensate 
wave function and its gradient gives the superfluid 
velocity \cite{sala-josephson,sala-new}. }

Under an external potential $U({\bf r})$ acting on individual atoms, 
the properties of the Fermi superfluid in the BCS-BEC crossover 
can be described, in the spirit of the density functional theory 
\cite{kohn1,kohn2,kohn}, by the energy functional 
\beq
E = \int \Big\{
{\hbar^2\over 2m_{p}}|\nabla \Psi({\bf r})|^2 +
U_p({\bf r}) |\Psi({\bf r})|^2
+ {\cal E}[n_p({\bf r})] \Big\} \ d^3{\bf r}
\label{energy-func}
\; ,
\eeq
where $U_{p}({\bf r})=2 U({\bf r})$ is the external
potential acting on a pair and $U({\bf r})$ 
the external potential acting on 
a single atom \cite{kim-zubarev,manini05,sala-josephson}. 
Here ${\cal E}[n_p]$ is bulk energy density as a function 
of density of pairs, i.e. the energy density 
of the uniform system in the BCS-BEC crossover 
\cite{kim-zubarev,manini05,sala-josephson}. 

The last two terms in  (\ref{energy-func}) correspond
to the local density approximation (LDA), equivalent to
hydrostatics, and the first term involving the gradient corresponds
to a correction to LDA \cite{kohn,rupak}. 
Including only the LDA term in 
energy density leads to the Thomas-Fermi (TF) approximation. 
In the gradient term $m_{p}=2m$ is the 
mass of a pair with $m$ the mass of one atom. This term takes into
account corrections to the kinetic energy due to spatial variations in
the density of the system. For normal fermions various
authors have proposed different gradient terms 
\cite{von,tosi,SKA1,zaremba,sala-gradient}. For superfluid fermions we 
are
using the familiar von Weizs\"acker \cite{von} term in the case of 
pairs. 

We notice that the present gradient term is exactly the one 
that emerges \cite{strinati} from the Bogoliubov-de Gennes equations 
in the BEC regime of the BCS-BEC crossover. In this 
case the energy functional (\ref{energy-func}) is nothing but the 
Gross-Pitaevskii energy functional of the Bose-condensed molecules 
\cite{strinati,dicastro}. 

In addition, we stress that the gradient term 
$[\hbar^2/(2m_p)]\left[{\nabla \sqrt{n_p({\bf r})}} \right]^2$ 
of  (\ref{energy-func}) is very close to  that 
recently obtained in  \cite{rupak} 
for the same gradient term 
in the case of a Fermi gas at unitarity, in a rigorous calculation using 
an $\epsilon$ expansion of energy density  
around $d=4-\epsilon$ spatial dimensions \cite{rupak,son}. 
Their final result for this term  $0.032 
\hbar^2[\nabla n({\bf r})]^2/[mn({\bf r})]$, quoted in their (53),  
can be rewritten as $0.512(\hbar^2/m_p)\left[{\nabla 
\sqrt{n_p({\bf r})}} \right]^2$, if we recall $m_p=2m$ and $n({\bf r})= 
2|\Psi({\bf r})|^2= 2n_p({\bf r})$ $-$ very close to the present 
gradient term in (\ref{energy-func}). 

In the BCS-BEC crossover,  the bulk equation of state of energy 
density ${\cal E}(n)$ as a function of density of atoms $n$ 
of a dilute superfluid depends on the s-wave scattering length $a$
of the inter-atomic potential \cite{kim-zubarev,manini05,sala-josephson}. 
In the unitarity limit, when $a\to \pm \infty$,
the bulk energy density is independent of $a$ and,
for simple dimensional reasons \cite{baker,cowell,hei,sala-fermi},
must be of the form 
\beq
{\cal E}(n) = {3\over 5} \chi {\hbar^2 \over m} n^{5/3}
=2^{2/3}{6\over 5}\chi{\hbar^2 \over m} n_p^{5/3}, 
\label{energy-unitarity} 
\eeq
where $\chi$ is a universal coefficient. Thus,
in the unitarity limit the bulk chemical potential is proportional to
that of a non-interacting Fermi gas.
For fermions, results of fixed-node Monte Carlo  calculation 
give $\chi = (3\pi^2)^{2/3}\xi/2$ with $\xi=0.44$ \cite{mc}. 

Taking into account  (\ref{energy-unitarity}),
the energy functional of the Fermi superfluid reads
\beq
E = \int \Big\{
{\hbar^2\over 4m}|\nabla \Psi({\bf
r})|^2 + 2U({\bf r}) |\Psi({\bf r})|^2 
+ 2^{{2/3}} {{6}\over 5} \chi
{\hbar^2\over m} |\Psi({\bf r})|^{10/3} \Big\}
\ d^3{\bf r} \; .
\label{energy}
\eeq
By minimizing (\ref{energy}) with  constraint (\ref{norma})
we obtain the following nonlinear Schr\"odinger equation
\beq
\left[ -{\hbar^2 \over 4m}\nabla^2 + 2 U({\bf r}) 
+ 2^{{2/3}}
 \chi
{{{2}}
\hbar^2 \over m} |\Psi({\bf r})|^{4/3} \right]
\Psi({\bf r}) = 2 \mu_0 \Psi({\bf r}) \, ,
\label{gle}
\eeq
where the chemical potential $\mu_0$ is fixed by normalization. 
This is the zero-temperature  3D NLS equation  
of superfluid Fermi gas at unitarity. 
 
In the case of a large number of fermions, in  
(\ref{gle}) the nonlinear term vastly dominates over the kinetic 
energy term containing the Laplacian $\nabla^2$, and the numerical 
solution of this equation is thus insensitive to the Laplacian term.  
In phenomenological treatment of Fermi superfluid, this term is often 
neglected and the resulting model is called the TF model (LDA approach),  
which, however, leads to a nonanalytic solution in space variable. The 
inclusion of the Laplacian term leads to a solution of  (\ref{gle})
analytic in space variable, although lying close to the solution 
of the TF model for a large number of particles. 
The Laplacian term, however,  plays  an important role in the case 
of a small number of particles, as we shall show 
studying the dimensional crossover of the system. 

\begin{figure}[tbp]
\begin{center}
{\includegraphics[width=\linewidth]{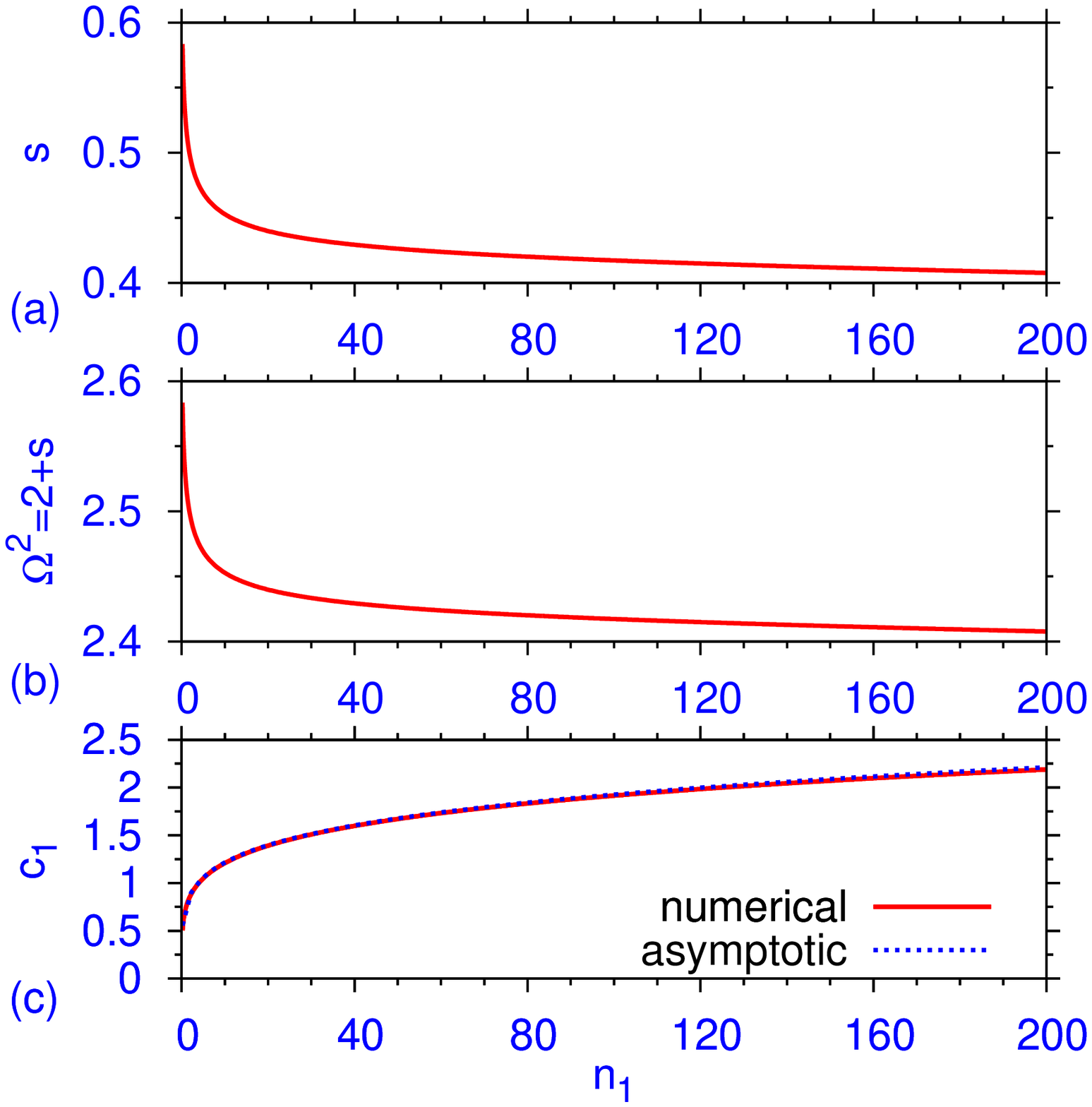}}
 \end{center} \caption{Ground-state energy $E$ 
(in units of $\hbar\omega$) versus $N$ from a solution of 
the 3D NLS equation (solid line). 
The results of Green-function Monte Carlo (GFMC)  \cite{FN1} 
and fixed-node diffusion Monte Carlo  (FNDMC)
\cite{FN} are shown for a comparison (symbols) 
The local density approximation (LDA), i.e. 
the Thomas-Fermi model, is also exhibited (dashed line).}
\label{lda}
\end{figure}

Now we compare the energies of our density functional (\ref{energy}) 
with those of Monte Carlo calculations \cite{FN,FN1} 
in the case of a harmonically trapped system with small number $N$ 
of spin 1/2 fermions at unitarity. We solve (\ref{gle}) 
numerically, using the Crank-Nicholson method detailed in Sec. \ref{V},  
and calculate the total energy $E$ of the system using  (\ref{energy}) 
for different $N$ in a spherically-symmetric harmonic trap 
\beq 
U(r)={1\over 2} m  \omega^2  r^2  
\eeq
with $\omega$ the trap frequency. 
In figure \ref{lda} we plot the total energy $E$ in units of 
$\hbar\omega$ vs. $N$. For comparison, we also plot the TF energy. 
The energy of the TF approximation (LDA approach) is 
analytically known to be $E(N)=(3N)^{4/3}\sqrt \xi/4$ \cite{bulgac}. 
{From the data of Figure \ref{lda} we find that the 
average percentage 
deviation  of the present result from the FNDMC data is 
3.30\%, whereas the same of the LDA result from the FNDMC data
is 9.47\%.} 
{Very recently it has been shown \cite{toigo} that 
the best fit to the fixed-node diffusion Monte-Carlo data at unitarity 
\cite{FN} is obtained, in the case of even number of particles, 
by using $\lambda=0.18$ in the gradient term 
$\lambda \hbar^2|\nabla \Psi|^2/m$ of the energy functional (\ref{energy}).}

The usefulness of the present model 
equation (\ref{gle}) over the commonly-used TF approximation 
(LDA approach) is well appreciated. In fact, 
the TF approximation is valid only for very large $N$ values when the 
gradient term is negligible compared to the bulk chemical potential. 
{The present density-functional model for Fermi 
superfluid at unitarity has  been 
extended over the full BCS-unitarity crossover and the extended 
model has been found to yield \cite{sadhan} energy of a Fermi superfluid 
under spherical harmonic confinement  
in good agreement with Monte Carlo 
data \cite{blume} in the crossover domain. }

\section{External confinement: axially-symmetric harmonic potential}
\label{III}

If we take ${\bf r}\equiv (\tilde \rho, \tilde z)$ so that $\tilde \rho$ 
and $\tilde z$ are the radial and axial variables and if we consider the  
axially-symmetric harmonic  trap 
\beq \label{pot}
U({\bf r})= \frac{1}{2}m \omega_\bot^2 (
\lambda_2^2\tilde \rho^2 +\lambda_1^2 \tilde 
z^2),
\eeq
where $\lambda_2\omega_\bot$ and $\lambda_1\omega_\bot$ are the trap 
frequencies in 
radial and axial directions 
with $\lambda_1$ and $\lambda_2$ convenient trap  parameters. 
In the following we shall consider for simplicity 
a real wave function $\Psi({\bf r})$. 
Equation (\ref{gle}) can be reduced to the 
following 
dimensionless form by scaling $\tilde z= z a_\bot$, $\tilde \rho =\rho 
a_\bot$, $\Psi= \psi\sqrt{N/2}/a_\bot^{3/2}$, 
and $\mu_0=\tilde \mu_0 
\hbar 
\omega_\bot$  
\beqa
\biggr[ -{1 \over 4}\biggr(\frac{\partial^2}{\partial  \rho^2} 
&+&\frac{\partial^2}{\partial  z^2}
\biggr)-\frac{1}{{4} \rho}\frac{\partial}{\partial  \rho}
+\lambda_2^2 \rho^2+\lambda_1^2 z^2\nonumber \\
&+& 2N^{{2/3}}\chi
  \psi^{4/3}  \biggr]
\psi(\rho,z) = 2 \tilde \mu_0 \psi(\rho,z) \, ,
\label{gle2}
\eeqa
where $a_\bot=\sqrt{\hbar/m\omega_\bot}$ 
is the characteristic harmonic oscillator length in the 
radial direction. Equation (\ref{gle2}) satisfies 
normalization $2\pi \int_0^\infty  d\rho \int_{-\infty}^\infty
dz \rho \psi^2(\rho,z)=1$. 

For a cigar-shaped superfluid it is convenient to define a linear 
density through
\beq\label{denl}
f^2(z)=2\pi \int_0^\infty  d\rho 
 \rho \psi^2(\rho,z).
\eeq  
Similarly, for a disk-shaped superfluid it is convenient to consider a 
radial density through  
\beq\label{denr}
\phi^2(\rho)= \int_{-\infty}^\infty  dz \psi^2(\rho,z).
\eeq

\section{Cigar-shaped Fermi superfluid: 3D-1D crossover}
\label{IV}

\subsection{Harmonic Confinement}

Now let us suppose that the external trapping potential $U({\bf r})$
is given by a harmonic confinement of
frequency $\omega_{\bot}$ in the cylindrical radial direction
$\tilde \rho$ and by a generic potential
$V(\tilde z)$ in the cylindrical axial direction $\tilde z$:
\beq
U({\bf r}) = {1\over 2} m \omega_{\bot}^2\tilde \rho^2 + 
V(\tilde z) \; .
\eeq
We introduce  the variational field
\beq
\Psi({\bf r}) = {1\over \pi^{1/2}\tilde  \sigma(\tilde z)}
\exp{\left(-{\tilde \rho^2\over 2 \tilde \sigma
^{{2}}
(\tilde z)}\right)} \ \tilde  f(\tilde z)
\eeq
into the fermionic energy functional (\ref{energy})
and integrate over $\tilde x$ and $\tilde y$ coordinates. 
After neglecting the space derivatives
of $\tilde \sigma(\tilde z)$ (adiabatic approximation)
we obtain the following effective energy functional
\beqa
E_1 = \int_{-\infty}^\infty \Big\{
{\hbar^2\over 4m}
\Big[ {d \tilde f(\tilde z)\over d \tilde z}
\Big]^2
+
\Big[ 2V(\tilde z)
+ {\hbar^2\over 4m \tilde \sigma
^{{2}}
(\tilde z)}
\nonumber
\\
+ m\omega_{\bot}^2 \tilde
\sigma
^{{2}}(\tilde z)
  \Big] \tilde f^2(\tilde z)
+ 2^{{2/3}} {{18}\chi \over 25
\pi^{2/3}}
{\hbar^2 \over m} {\tilde f^{10/3}(\tilde z) \over\tilde \sigma
^{{4/3}}
(\tilde z)}
\Big\}
\ d\tilde z \; ,
\label{effective1d}
\eeqa
which depends on two fields: the transverse width $\tilde \sigma(\tilde z)$
and the axial wave function $\tilde f(\tilde z)$. 

Note that the variational approach we are using here 
has been successfully 
applied in the dimensional reduction from 3D to 1D 
of the 3D Gross-Pitaevskii equation, 
which describes trapped Bose-Einstein condensates.  
In that case the variational approach 
ends up with the 1D nonpolynomial Schr\"{o}dinger
equation (1D NPSE) for the axial wave function \cite{sala-npse}. 
The 1D NPSE has been extended to investigate the Tonks-Girardeau 
regime \cite{sala-tonks(a),sala-tonks(b)}, the two-component 
BEC \cite{sala-npse-2cp}, 
transverse spatial modulations \cite{sala-npse-tr} and also axial 
vorticity \cite{sala-npse-vor}, 

Minimizing ${ E}_1$ with respect to $\tilde f(\tilde z)$ one finds
\beqa
\Big[ -{\hbar^2\over 4m} {d^2\over d \tilde z^2} + 2 V(\tilde z)
+ {\hbar^2\over 4m \tilde \sigma
^{{2}}(\tilde z)
} + m\omega_{\bot}^2  \tilde \sigma
^{{2}}
(\tilde z)
\nonumber
\\
+ 2^{{2/3}}
{{{6}}
\chi \over 5 \pi^{2/3}}
{\hbar^2 \over m}
{\tilde f^{4/3}(\tilde z) \over \tilde \sigma
^{{4/3}}
(\tilde z)} \Big]\tilde  
f(\tilde z) = 2\tilde  \mu_1 \ \tilde f(\tilde z) \; .
\label{npse1d}
\eeqa
This is a 1D Schr\"odinger equation 
and $\tilde \mu_1$ is fixed by the normalization
\beq \label{nr1}
\int_{-\infty}^\infty \tilde f^2(\tilde z) \ d\tilde z = {N\over 2}=N_p \; .
\eeq
Instead, minimizing ${ E}_1$
with respect to $\tilde \sigma$ one gets
\beq
a_{\bot}^{-4} \tilde \sigma^{
4}(\tilde z) = {1\over 4} + 2^{
{2/3}} {{{12}}\chi
\over 25 \pi^{2/3}} \tilde f^{4/3}(\tilde z) \tilde \sigma^{{2/3}}(\tilde z)
 \; .
\label{al1d}
\eeq
We call  (\ref{npse1d}), equipped with (\ref{al1d}), 
the 1D nonpolynomial Schr\"odinger (1D NPS) equation. 

The 1D NPS equation can be 
conveniently written in dimensionless form by scaling
$\tilde z= z a_\bot$,
$\tilde \sigma(\tilde z)= \sigma( z) a_\bot$, 
$\tilde f(\tilde z)= \sqrt {N/2} f( z)/\sqrt a_\bot$,
and 
$\tilde \mu_1=\hbar \omega_\bot ( \mu_1+1/2)$ as follows:
\beqa
\Big[ -{1\over 4} {d^2\over d z^2} +\lambda_1^2 z^2
+ {1\over 4  \sigma
^{{2}}( z)
} +  \sigma
^{{2}}
( z)-1
\nonumber
\\
+ N^{{2/3}}
{{{6}}
\chi \over 5 \pi^{2/3}}
{ f^{4/3}( z) \over  \sigma
^{{4/3}}
( z)} \Big]  f( z) = 2  \mu_1  f( z) \; ,
\label{npse1d1}
\eeqa
\beq
  \sigma^{
4}( z) = {1\over 4} + N^{
{2/3}} {{{12}}\chi
\over 25 \pi^{2/3}}  f^{4/3}( z)
\sigma^{{2/3}}( z)
 \; ,
\label{al1d1}
\eeq
with normalization $
\int_{-\infty}^\infty f^2( z) \ d z = 1 \; .$
Here we have included a constant term [=1/2, corresponding to the 
energy of the system in the transverse trap whose effect has been 
integrated out in  (\ref{npse1d1}) and (\ref{al1d1})] 
in the scaled chemical 
potential, so that in the $N=0$ limit the chemical potential coincides 
with that of the axial harmonic trap. In the $N=0$ limit, from  
(\ref{al1d1}) we find  that $\sigma^2(z)=1/2$ and 
the $1/(4\sigma^2)+\sigma^2$ terms cancel the constant term in  
(\ref{npse1d1}).  
In deriving  (\ref{npse1d1}) we assumed an harmonic confinement in
the $z$ direction: $V(\tilde z)=m\lambda_1^2 \omega_\bot^2 
\tilde z^2/2$
with $\lambda_1$ the anisotropic parameter.

By using the 1D NPS equation (\ref{npse1d1}) with (\ref{al1d1})
we can study the dimensional crossover from 3D to 1D of 
the superfluid Fermi gas at unitarity. In general  (\ref{al1d1}) 
must be solved numerically for $\sigma(z)$ in terms of 
$f(z)$ and the result when substituted in  (\ref{npse1d1}) leads
to the principal result of this section. No closed-form analytic
expression for this equation can be given as in the bosonic case
\cite{sala-npse}, except under special limiting conditions. 

The first interesting limit of the formalism is obtained for small 
number of atoms when the 
nonlinear term in  (\ref{npse1d1}) is  small, (which
corresponds to the weak-coupling limit in the bosonic case,) so that the
last term in  (\ref{al1d1}) can be neglected and the transverse width
is $z$ independent and is given by  
$
\sigma( z) \equiv  \sigma =
1/\sqrt{2}$ under the condition
$N_pf^2( z) \ll
125 \sqrt{2}\pi/({
{128}}(3\chi)^{3/2})={0.273293}...$ 
(obviously satisfied for a small number of fermions). The
cigar-shaped system is then quasi-1D and governed by
\beqa
\Big[ -{1\over 4} {d^2\over d z^2} +\lambda_1^2 z^2
+
{{{6}}N_p^{2/3}
\chi (2 f)^{4/3}
\over 5 \pi^{2/3}}
 \Big]  f( z) = 2  \mu_1 \
f(
z) \; .
\label{npse1d2}
\eeqa

In the opposite extreme, for a large number of fermions,  $N_p f^2( z) \gg
125 \sqrt{2}\pi/({{128}}(3 \chi)^{3/2})$,
the cigar-shaped system is effectively 3D. Under this condition
 (\ref{al1d1}) can also be solved for $ \sigma( z)$ to yield
$ \sigma( z) =[4 \sqrt N_p f( z)/\sqrt \pi]^{2/5}
(3\chi/25)^{3/10}$. Consequently,  the
$1/(4 \sigma^2( z))$ term  in  (\ref{npse1d1}) can be
neglected and the remaining terms combined to yield
\beqa
\Big[ -{1\over 4} {d^2\over d z^2} +
\lambda_1^2 z^2
+\frac{7}{5}\frac{(6\chi)^{3/5}N_p^{2/5} f^{4/5}}{(5\pi^2)^{1/5}}
 \Big]  f( z) = 2  \mu_1 \
f(
z) \; . \nonumber \\
\label{npse1d3}
\eeqa

The power of the nonlinear term has changed from 7/3 to 9/5 as we pass
from quasi-1D regime governed by  (\ref{npse1d2}) to the 
effectively 3D regime
governed by  (\ref{npse1d3}). In the quasi-1D regime the nonlinear
power 7/3 is the same as that in the original three-dimensional equation
(\ref{gle}), whereas in the 3D regime it has acquired a different power.
Although  the quasi-1D equation (\ref{npse1d2}) has been used in
different studies on a Fermi superfluid \cite{adhi} and on a degenerate 
Fermi gas \cite{adhi2},  (\ref{npse1d3}) for Fermi superfluid
is new. We shall see that  (\ref{npse1d3}) is already valid for a 
moderate number of Fermi atoms {($N>100$)} and is of 
interest in 
phenomenological applications. 

\begin{figure}[tbp]
\begin{center}
{\includegraphics[width=.9\linewidth]{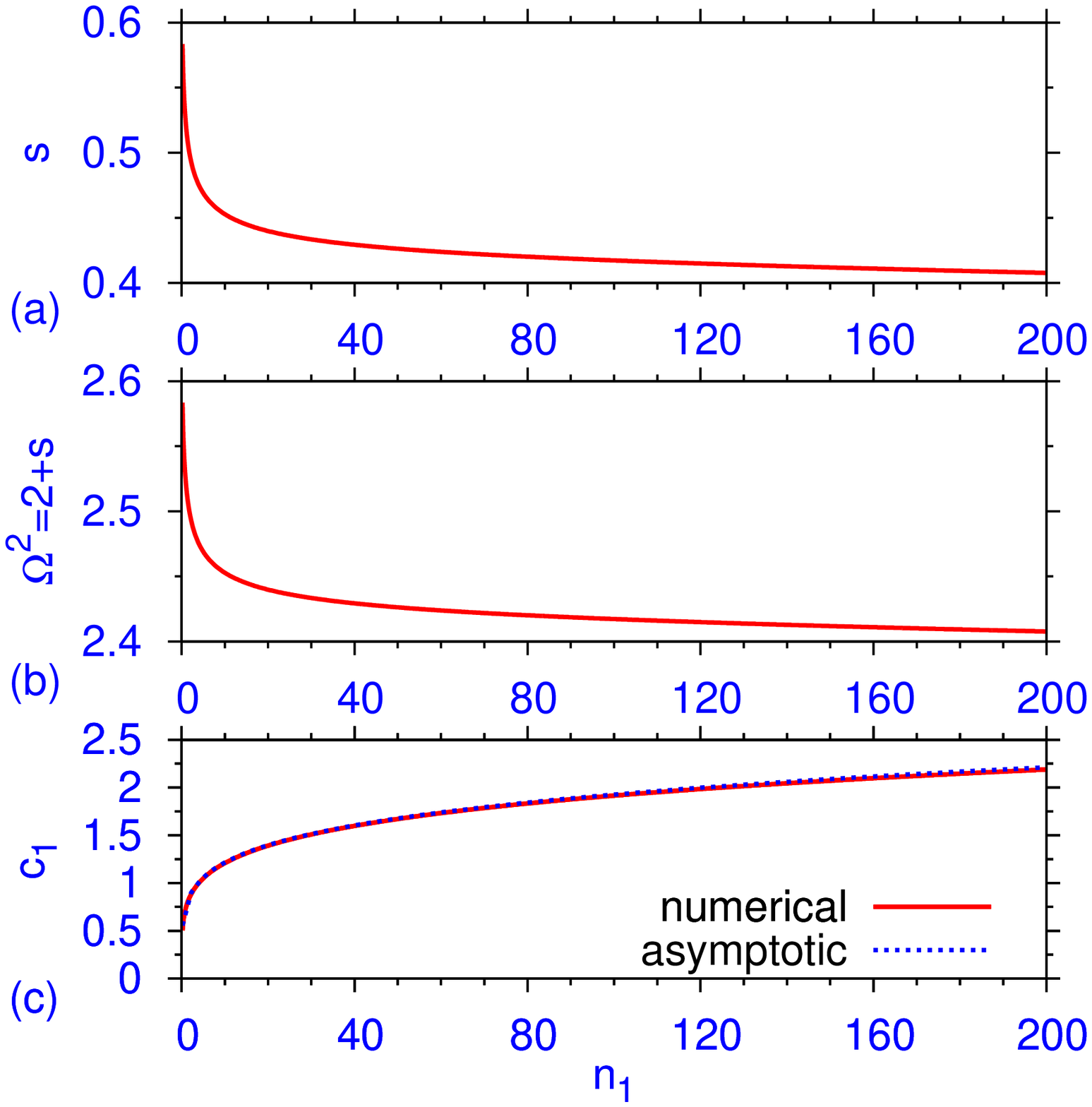}}
 \end{center}
\caption{(a) The $s$ vs. $n_1$ dependence as calculated 
from a 
numerical solution of  (\ref{cp1}) and (\ref{cp2}). 
 (b) Square of the frequency  
$\Omega^2= 2+s$ in a cigar-shaped trap vs. 
density $n_1$ as calculated from  (\ref{cz2}). 
(c) Sound 
velocity $c_1$ vs. density $n_1$ of a uniform gas from a numerical 
solution of  (\ref{mu1-a}) and (\ref{mu1-b}) and the asymptotic 
$c_1 ={0.766846} n_1^{1/5} $
result valid for large $n_1$. All results refer to the unitarity limit 
$\xi=0.44$. }
\label{fg1}
\end{figure}

In the 3D regime it is a good approximation to neglect the kinetic
energy term in  (\ref{npse1d3}) 
and the following analytic expression for density is obtained
in the so called TF approximation
\beqa
N_p  f^2( z)&=& \frac{ 125 \pi}{7(42\chi)^{3/2}}
(2 \mu_1-\lambda_1^2 z^2)^{5/2}\Theta(2
\mu_1-\lambda_1^2 z^2 ),\nonumber  \\
&\approx & \frac{0.206105}{\chi^{3/2}} (2 \mu_1-\lambda_1^2
z^2)^{5/2}
\Theta(2\mu_1-\lambda_1^2 z^2 )
\label{TF1}
\eeqa
where $\Theta(x)$ is the Heaviside step function. As we are in the 3D regime
it is interesting to compare  this result with the following TF
approximation made
on the full three-dimensional equation (\ref{gle}) after integrating
over the transverse variables
\beqa
 N_p  f^2( z)&=& \frac{ \pi}{10 \sqrt 2\chi^{3/2}}
(2 \mu_1-\lambda_1^2 z^2)^{5/2}
\Theta(2\mu_1-\lambda_1^2 z^2 ),\nonumber  \\
&\approx & \frac{0.222144}{\chi^{3/2}} (2 \mu_1-\lambda_1^2
z^2)^{5/2}
\Theta(2 \mu_1-\lambda_1^2 z^2 ).
\label{TF2}
\eeqa
The two TF results have the same functional dependence on the variables
as well as very similar numerical coefficients, in spite of
 (\ref{npse1d3}) and (\ref{gle}) having different powers of
density in the nonlinear terms. The quasi-1D equation 
(\ref{npse1d2}) has the same power of density in the nonlinear term as
 (\ref{gle}). Nevertheless, a TF approximation made in 
(\ref{npse1d2}) will generate a density with an entirely
different dependence on $ z$.

It is now interesting to study the bulk chemical potential
$\mu_1(n_1)$ implicit in   
(\ref{npse1d1}) and (\ref{al1d1}) given by
\beqa
\mu_1(n_1)= {1\over 4  \sigma
^{{2}}( z)
} +  \sigma
^{{2}}
( z)-1
+ 2^{{2/3}}
{{{6}}
\chi \over 5 \pi^{2/3}}
{ n_1^{2/3}( z) \over  \sigma
^{{4/3}}
( z)}
\label{cp1}\\
  \sigma^{
4}( z) = {1\over 4} + 2^{
{2/3}} {{{12}}\chi
\over 25 \pi^{2/3}}  n_1^{2/3}( z)
\sigma^{{2/3}}( z)
 \; ,
\label{cp2}
\eeqa
where density $n_1( z)= N_p {f^2(z)}$.
Many physical observables are determined by the bulk chemical
potential so obtained.
For example, assuming a power-law dependence
$\mu_1 \sim \; n_1^{s}$
for the bulk chemical potential on density $n_1$ (polytropic equation of
state), the frequency of the lowest axial compressional mode 
{$\Omega_1$}is given by
\cite{sala-tonks(a)} 
\beq   \label{cz2}
\Omega^2 \equiv \left[\frac{\Omega_1}{\lambda_1 \omega_\bot}\right]^2 = 
2 + s   \; .
\eeq 
Here we introduce an effective polytropic index
$s$ as the logarithmic derivative of the bulk chemical
potential $\mu_{{1}}$, that is
\beq
s = {n_1 \over \mu_1} {\partial \mu_1 \over \partial n_1} \; .
\eeq
From  (\ref{cp1}) and (\ref{cp2}) one finds 
that in the 1D regime $s=2/3$ and 
$\Omega=\Omega_1/(\lambda_1\omega_\bot) =\sqrt{8/3}$, 
while in the 3D regime $s=2/5$ and 
$\Omega=\Omega_1/(\lambda_1 \omega_\bot)=\sqrt{12/5}$. 
Note that the result $\sqrt{12/5}$ is exactly the same 
one obtains setting $\gamma =2/3$ in the formula 
$(3\gamma+2)/(\gamma +1)$ derived by Cozzini and Stringari 
\cite{coz} for the axial breathing mode of 
3D cigar-shaped superfluids with a 3D bulk chemical potential 
that scales as $n_p^{\gamma}$. 

In figure \ref{fg1} (a) we show the power 
$s$ of the dependence  $\mu_1 \sim n_1^s$ as a function of $n_1$ 
obtained by numerically solving nonlinear  (\ref{cp1}) and 
(\ref{cp2}). 
In figure \ref{fg1} (b)  we plot the square of collective
frequency $\Omega^2$ in the dimensional crossover as a function
of the axial density $n_1$ and 
As the number of atoms $N$ increases, the frequency of 
axial compressional mode slightly decreases with a decrease 
of the polytropic power in $\mu \sim n_1^s$ relation. 

\subsection{Uniform Density}

Now we consider the axially uniform case, where $V(z)=0$.
In this case all the variables attain a constant value
independent of $z$ and we remove the $z$ dependence on all variables.
Setting $n_1=N_p  f^2$, from  (\ref{npse1d1})
and (\ref{al1d1}) we find
\beqa
 \mu_1 &=& {1\over 8  \sigma^2
} + {1\over 2}  \sigma^2-\frac{1}{2}
+ {3\chi \over 5 }
{{ \left[
{2n_1 \over \pi
 \sigma^2}
\right]^{2/3}  }}\; ,
\label{mu1-a}
\\
&&   \sigma^4 = {1\over 4} +
{12\chi    \over 25 }
{{\left[ \frac{2n_1 \sigma}{\pi} \right]^{2/3}}}
\; .
\label{mu1-b}
\eeqa
These equations can be used to derive
the axial sound velocity $c_1$ of the fermionic system,
that is obtained with the formula \cite{landau2}
\beq
c_1 = \sqrt{{n_1} {\partial \mu_1 \over \partial n_1}} \; .
\eeq
In the quasi-1D regime (for small $n_1$), from  (\ref{mu1-a}) and 
(\ref{mu1-b})
we obtain $ \sigma^2=1/2,$ $ \mu_1=
3\chi(4n_1/\pi)^{2/3}/5$
and $
c_1= \sqrt{ 2\chi/5}(4n_1/\pi)^{1/3}
$. In the effective 3D regime (for large $n_2$), from  (\ref{mu1-a}) 
and (\ref{mu1-b})
 we obtain $\sigma^2=(12 \chi/25)^{3/5}(2n_1/\pi)^{2/5}, \mu_1=(7/5)
(3\chi)^{3/5} (n_1/\pi)^{2/5}/
20^{1/5}$ and $c_1=\sqrt{14/25}(3\chi)^{3/10} (n_1/\pi)^{2/10}/
20^{1/10} ={0.766846} n_1^{1/5}  $ for $\xi=0.44$.
In figure \ref{fg1} (c) we plot the sound velocity
$c_1$ as a function of the axial density $n_1$ as calculated from a full 
numerical solution of  
(\ref{mu1-a}) and (\ref{mu1-b}) as well as the asymptotic result $c_1 
={0.766846}  n_1^{1/5}  $ valid for the effective 3D 
regime for large 
$n_1$. The two results are indistinguishable except near the origin. 
This shows that the effective 3D description (\ref{npse1d3})  is very 
good 
except for very small atom number.    The sound velocity increases with 
matter density as it should.  

\section{Disk-shaped Fermi superfluid: 3D-2D crossover}
\label{V}

Let us suppose that the external trapping potential $U({\bf r})$
is given by a generic potential $W(\tilde \rho)$ in the cylindrical
radial direction $\tilde \rho$ and by a
harmonic confinement of frequency $\omega_z$
in the cylindrical axial direction $\tilde z$:
\beq
U({\bf r}) = W(\tilde \rho ) + {1\over 2} m \omega_z^2 \tilde 
z^2 \; .
\eeq
We introduce the variational field
\beq
\Psi({\bf r}) = {1\over \pi^{1/4} \tilde \eta^{{1/2}}
(\tilde \rho)}
\exp{\left(-{z^2\over 2\tilde \eta^{{2}}
(\tilde \rho )}\right)} \tilde  \phi(\tilde \rho )
\eeq
into the fermionic energy functional (\ref{energy})
and integrate over the $\tilde z$ coordinate.
After neglecting the space derivatives
of $\tilde \eta(\tilde \rho )$ we obtain the following effective energy
functional
\beqa
E_2 &=& 2\pi
\int_0^\infty \Big\{
 {\hbar^2\over 4m}
[\nabla_{\tilde \rho}\tilde \phi(\tilde \rho)]^2   +
\Big[
2 W(\tilde \rho )
+ {\hbar^2\over 8 m \tilde \eta^{{2}}
(\tilde \rho )}
\nonumber
\\
&+&
{1\over 2} m\omega_{\tilde z}^2 \tilde \eta^{{2}}
(\tilde \rho )
\Big]
\tilde \phi^2(\tilde \rho )
+ 2^{2/3} {2\chi \over \pi^{1/3}} \left({3\over 5}\right)^{3/2}
\nonumber
\\
&\times&
{\hbar^2 \over m} {\tilde \phi^{10/3}(\tilde \rho) \over
\tilde \eta ^{{2/3}}(\tilde \rho )}
\Big\} \  \tilde  \rho \ d\tilde \rho \; ,
\label{effective2d}
\eeqa
which depends on two fields:
the axial width $\tilde \eta(\tilde \rho )$ and the transverse wave
function
$\tilde \phi(\tilde \rho )$. 

Also in this case we observe that the variational approach we are using 
has been successfully applied in the dimensional reduction from 3D to 2D 
of the 3D Gross-Pitaevskii equation. The resulting 
effective equation has been called 2D nonpolynomial 
Schr\"odinger equation (2D NPSE) \cite{sala-npse}. 

Minimizing $E_2$ with respect to $\tilde \phi(\tilde \rho
)$
one finds
\beqa
\Big[ -{\hbar^2\over 4m} \nabla_{\tilde\rho}^2 + W(\tilde\rho)
+ {\hbar^2\over 8m \tilde\eta^{{2}}
(\tilde\rho )}
+ {1\over 2} m\omega_z^2 \tilde\eta^{{2}}
(\tilde\rho )
\nonumber
\\
+2^{2/3} {\sqrt{3\over 5}}{2\chi \over  \pi^{1/3}} {\hbar^2
\over m} {\tilde\phi^{4/3}(\tilde\rho) \over \tilde\eta^{{2/3}}
(\tilde\rho ) }
\Big] \tilde\phi(\tilde\rho ) = 2 \tilde \mu_2 \tilde\phi(\tilde\rho )
\; .
\label{npse4d}
\eeqa
This equation  is a two-dimensional Schr\"odinger equation
and $\tilde \mu_2$ is fixed by the normalization
\beq
2 \pi
\int_0^\infty \tilde\phi^2(\tilde\rho) \
\ \tilde\rho \ d\tilde\rho = {N\over 2}
\; .
\eeq
Instead, minimizing $E_2$ with respect to $\tilde\eta(\tilde\rho)$ one
gets
\beq
a_z^{-4} \tilde\eta^4(\tilde\rho ) = {1\over 4}  + 2^{2/3} {4 \chi \over
3
\pi^{1/3}}
\left({3\over 5}\right)^{3/2} \tilde\phi^{4/3}(\tilde\rho)
\tilde\eta^{4/3}
(\rho)
\; ,
\label{al4d}
\eeq
where $a_z=\sqrt{\hbar/(m\omega_z)}$ is the
characteristic harmonic length
in the axial direction. 
We call  (\ref{npse4d}), together with  (\ref{al4d}), 
the 2D nonpolynomial Schr\"odinger (2D NPS) equation. 

The 2D NPS equation can be
conveniently written in dimensionless form by scaling
$\tilde \rho=\rho a_z$,
$\tilde \eta(\tilde \rho )= \eta( \rho) a_z$, $\tilde \phi(\tilde
\rho)=\sqrt N_p
 \phi( \rho)/
a_z$,
and
$\tilde \mu_2=\hbar \omega_z  (\mu_2+1/4)$ as follows:
\beqa
\Big[ -{1\over 4} \nabla_{ \rho}^2 + { \rho^2}{\lambda_ 2^2}
+ {1\over 8 \eta^{{2}}
(\rho )}
+ {1\over 2}   \eta^{{2}}
( \rho )-\frac{1}{2}
\nonumber
\\
+N^{2/3} {\sqrt{3\over 5}}{2\chi \over  \pi^{1/3}}  { \phi^{4/3}
( \rho) \over  \eta^{{2/3}}
( \rho ) }
\Big]  \phi( \rho ) = 2  \mu_2  \phi( \rho ) \; ,
\label{npse5d}\\
 \eta^4( \rho ) = {1\over 4}  + N^{2/3} {4 \chi \over 3
\pi^{1/3}}
\left({3\over 5}\right)^{3/2}  \phi^{4/3}( \rho)  \eta^{4/3} ( \rho)
\; ,
\label{al5d}
\eeqa
with normalization $2\pi \int_0^\infty \phi^2(\rho)\rho d \rho =1$.
In deriving 
(\ref{npse5d}) and (\ref{al5d}) we assumed an harmonic confinement
in the radial $\tilde\rho$ direction:
$W(\tilde\rho)=m\omega_z^2\tilde \lambda_2^2\rho^2/2$. Again in defining 
the reduced chemical potential $\mu _2$ we have removed the zero-point 
energy corresponding to the energy of the axial trap, so that in the 
$N=0$ limit, 
(\ref{npse5d}) and (\ref{al5d}) coincides with  the corresponding linear 
harmonic oscillator problem.   

By using  (\ref{npse5d}) with (\ref{al5d})
we can study the dimensional crossover from 3D to 2D of
the superfluid Fermi gas at unitarity. First  (\ref{al5d})
is to be solved numerically for $\eta(\rho)$  in terms of $\phi(\rho)$
and the result when substituted in  (\ref{npse5d}) gives the desired
result for studying a crossover from 3D to 2D. Closed-form analytic
result for these equations is possible only under limiting conditions.

The first interesting limit of the formulation is obtained for a 
small number of atoms  when the
nonlinear term in  (\ref{npse5d}) is small, so that the last term in
 (\ref{al5d}) can be neglected. Under this condition $N_p \phi^2 \ll
25\sqrt{3\pi}(5/3)^{1/4}/
(192 \chi^{3/2})={0.148656}...$ (obviously satisfied for 
a small number of 
fermions)
the
longitudinal width $\eta(\rho)=\sqrt{1/2}$ is independent of $\rho$. The
disk-shaped system is then quasi-2D and described by
\beqa
\Big[ -{ \nabla_\rho^2 \over 4} + { \rho^2}{\lambda_2^2}
+ \sqrt{3\over 5}{4\chi N_p^{2/3} \over  \pi^{1/3}}  \phi^{4/3}
\Big]  \phi( \rho ) = 2  \mu_2  \phi( \rho ) \; ,\nonumber \\
\label{npse6d}
\eeqa

In the opposite extreme, for a large number of fermions, 
$N_p \phi^2 \gg 25\sqrt{3\pi}(5/3)^{1/4}/(192\chi^{3/2})$,
the disk-shaped system is effectively 3D. Under this condition
 (\ref{al5d}) can be solved for $\eta(\rho)$ to yield
$\eta(\rho)=2N_p^{1/4}\chi^{3/8}3^{3/16}\phi^{1/2}(\rho)/(5^{9/16}\pi^{1/8})$.
Substituting this result in  (\ref{npse5d}) and neglecting the
$1/[8 \eta^2(\rho)]$ term we get
\beqa
\Big[ -{ \nabla_\rho^2 \over 4} + { \rho^2}{\lambda_2^2}
+ \frac{12 \chi^{3/4}3^{3/8}\sqrt N_p \phi}{5\pi^{1/4}5^{1/8}}
\Big]  \phi( \rho ) = 2  \mu_2  \phi( \rho ) \; .
\label{npse7d}
\eeqa

\begin{figure}[tbp]
\begin{center}
{\includegraphics[width=\linewidth]{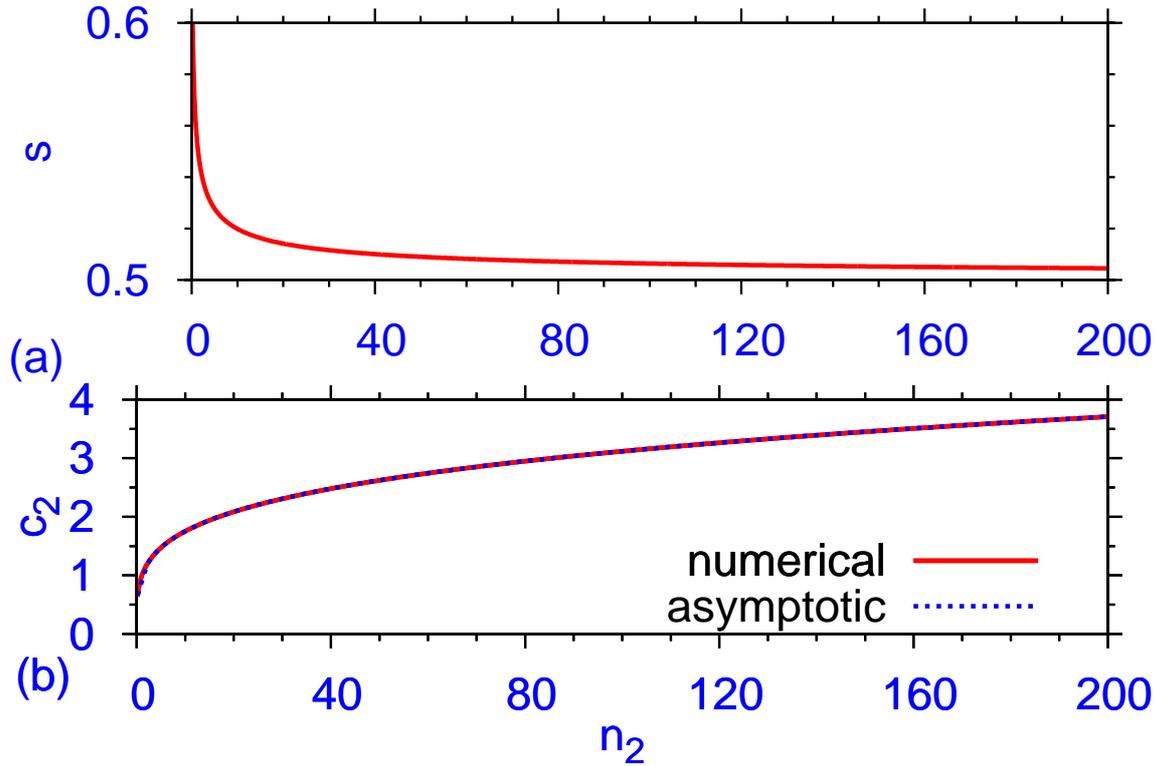}}
 \end{center}
\caption{(a) The  $s$ vs. $n_2$ dependence 
as calculated from a numerical solution 
of  (\ref{mu3}) $-$ (\ref{mu5}). 
(b) Sound
velocity $c_2$ vs. density $n_2$ of a uniform gas from a numerical
solution of  (\ref{mu1-c}) and (\ref{mu1-d}) and the 
asymptotic result $c_2={0.986212} 
n_2^{1/2}$ for large $n_2$.
All results refer to the unitarity limit
$\xi=0.44$.}
\label{fg2}
\end{figure}

The power of the nonlinear term has changed from 7/3 to 2 as we pass
from quasi-2D regime governed by  (\ref{npse6d}) to the 3D regime 
governed by  (\ref{npse7d}). In the quasi-2D regime the nonlinear
power 7/3 is the same as in the original three-dimensional equation
(\ref{gle}), whereas in the 3D regime it has acquired a different
power. Of the models (\ref{npse6d}) and (\ref{npse7d}),  
(\ref{npse6d}) was previously considered by others \cite{new}, whereas  
(\ref{npse7d}) is new. However, we shall see that  (\ref{npse7d}) 
should have wide phenomenological application for disk-shaped superfluid 
as this form is already effective for {$N>100$,} 
producing better 
approximation than (\ref{npse6d}).
  
In the 3D regime if we  neglect the kinetic
energy term in  (\ref{npse7d})
and the following analytic expression for density
is obtained in the TF approximation after a neglect of the kinetic 
energy term 
\beqa
N_p  \phi^2( \rho)&=& \frac{ 25 \sqrt\pi 5^{1/4}}{144\chi^{3/2}3^{3/4}}
\left(2 \mu_2-{\rho^2} {\lambda^2}\right)^{2}
\Theta(2\mu_2-{\rho^2}{\lambda^2}  ),\nonumber  \\
&\approx &
\frac{0.201862}{\chi^{3/2}}
\left(2 \mu_2-{\rho^2} {\lambda^2}\right)^{2}
\Theta(2\mu_2-{\rho^2}{\lambda^2}  ).
\label{TF3}
\eeqa
As we are in the 3D regime
it is interesting to compare  this result with the following TF
approximation made 
on the full three-dimensional equation (\ref{gle}) after integrating
over the longitudinal variable
\beqa
N_p  \phi^2( \rho)&=& \frac{3\pi}{32\chi^{3/2}\sqrt 2 }
\left(2 \mu_2-{\rho^2} {\lambda^2}\right)^{2}
\Theta(2 \mu_2-{\rho^2}{\lambda^2}  ),\nonumber  \\
&\approx &
\frac{0.20826}{\chi^{3/2}}
\left(2 \mu_2-{\rho^2} {\lambda^2}\right)^{2}
\Theta(2 \mu_2-{\rho^2}{\lambda^2}  ).
\label{TF4}
\eeqa
The two TF results have the same functional dependence on the variables
as well as very similar numerical coefficients, in spite of
 (\ref{npse7d}) and (\ref{gle}) having different powers of
density in the nonlinear terms. The quasi-2D equation
(\ref{npse6d}) has the same power of density in the nonlinear term as
 (\ref{gle}). Nevertheless, a TF approximation made in 
(\ref{npse6d}) will generate a density with an entirely
different dependence on $ \rho$.

We now consider the bulk chemical potential implicit in  (\ref{npse5d}) 
and (\ref{al5d})
\beqa
\mu_2(n_2)=
 {1\over 8 \eta^{{2}}
(\rho )}
+ {1\over 2}   \eta^{{2}}
( \rho )-\frac{1}{2}
+2^{2/3} {\sqrt{3\over 5}}{2\chi \over  \pi^{1/3}}  { n_2^{2/3}
( \rho) \over  \eta^{{2/3}}
( \rho ) } \; ,
\label{mu3}\\
 \eta^4( \rho ) = {1\over 4}  + 2^{2/3} {4 \chi \over 3
\pi^{1/3}}
\left({3\over 5}\right)^{3/2}  n_2^{2/3}( \rho)  \eta^{4/3} ( \rho)
\; ,
\label{mu4}
\eeqa
where density  $n_2(\rho)=N_p \phi^2(\rho)$. 
Again the bulk chemical potential can be considered to possess a 
power-law dependence on density: $\mu\sim n_2^s$, where the numerical 
coefficient $s$ can be extracted from a numerical solution of 
nonlinear  
(\ref{mu3}) and (\ref{mu4}) using the relation
\beq\label{mu5}
s=\frac{n_2}{\mu_2} \frac{\partial \mu_2}{\partial n_2}.
\eeq
The coefficient $s$ is of interest in 
the study of physical observables of interest. 
In the quasi-2D regime $s=2/3$ whereas in the 3D regime $s=1/2$. 
However, in the 
quasi-2D to 3D crossover the $\mu(n_2)$ dependence
is to be calculated numerically using  (\ref{mu3}) 
and (\ref{mu4}) and then the polytropic index$s$
can be calculated as a function of $n_2$ 
for the dimensional crossover.  In figure \ref{fg2} (a) 
we plot the polytropic power $s$ vs. density. 

\subsection{Uniform Density}

Now we consider the radially uniform case, where $W(\rho)=0$.
In this case all the variables attain a constant value
independent of $\rho$ and we remove the $\rho$ dependence on all 
variables.
Setting $n_2=N_p  \phi^2$, from  (\ref{npse5d})
and (\ref{al5d}) we find
\beqa
 \mu_2 &=& {1\over 16  \eta^2
} + {1\over 4}  \eta^2-\frac{1}{4}
+ \sqrt{3 \over 5 } \frac{\chi}{\pi^{1/3}}
{{ \left[
{2n_2 \over 
 \eta}
\right]^{2/3}  }}\; ,
\label{mu1-c}
\\
   \eta^4 &=& {1\over 4} +\frac{4\chi}{3\pi^{1/3}}\left(\frac{3}{5} 
\right)^{3/2} (2n_2\eta^2)^{2/3}
\; .
\label{mu1-d}
\eeqa
These equations can be used to derive
the {radial} sound velocity $c_2$ of the fermionic 
system,
that is obtained with the formula \cite{landau2}
\beq
c_2 = \sqrt{{n_2} {\partial \mu_2 \over \partial n_2}} \; .
\eeq
In the quasi-2D regime (for small $n_2$), from  
{(\ref{mu1-c}) and 
(\ref{mu1-d})}
we obtain $ \eta^2=1/2,$ $ \mu_2= 2\chi 
\sqrt{3/5}n_2^{2/3}/\pi^{1/3}$
and $
c_2= 2\sqrt{\chi/\sqrt{15}}
 n_2^{1/3}/\pi^{1/6}$. In the effective 3D regime (for large $n_2$), 
from  
{(\ref{mu1-a}) and 
(\ref{mu1-b})}
 we obtain $\eta^2= 4(\sqrt 3\chi)^{3/4} n_2^{1/2}/(5\pi^{1/4}5^{1/8}) 
, \mu_2=(6/5) (27/5)^{1/8}\chi^{3/4}n_2^{1/2}/\pi^{1/4}
$ and $c_2=\sqrt{3/5} (27/5)^{1/16} \chi^{3/8}  n_2^{1/2}/\pi^{1/8} = 
{0.986212} n_2^{1/2} $ for $\xi = 0.44$. 
In figure \ref{fg2} (b) we plot the sound velocity
$c_2$ vs.  the axial density $n_2$ as calculated from a full 
numerical solution of  
(\ref{mu1-c}) and (\ref{mu1-d}) as well as the asymptotic result 
$c_2={0.986212} n_2^{1/2}$ for large $n_2$ in the 
effective 3D 
description given by  (\ref{npse7d}). The asymptotic result is 
indistinguishable 
from the exact result except near very small $n_2$. This shows that the 
effective 3D description of the system is very good. 
 The sound velocity increases with 
matter density as it should.  

\section{Numerical Result}
\label{VI}

Next we study the effectiveness of the dimensional reduction of the 
3D GL equation (\ref{gle}) to 1D and 2D forms with a variation of
the number of fermions for cigar- and disk-shaped configurations.
We numerically solve the full 
3D equation as well as various 1D and 2D reduced equations by 
discretizing them by the semi-implicit Crank-Nicholson algorithm with 
imaginary time propagation \cite{num,sala-numerics,cn}. 
For numerical convenience we transform the 
chemical potential term $\mu \psi$ in the nonlinear equations to a 
time-dependent term $id\psi/dt$. The space and time steps used in 
discretization were typically 0.05 and 0.001 respectively. 

\subsection{3D-1D crossover}

\begin{figure}[tbp]
\begin{center}
{\includegraphics[width=\linewidth]{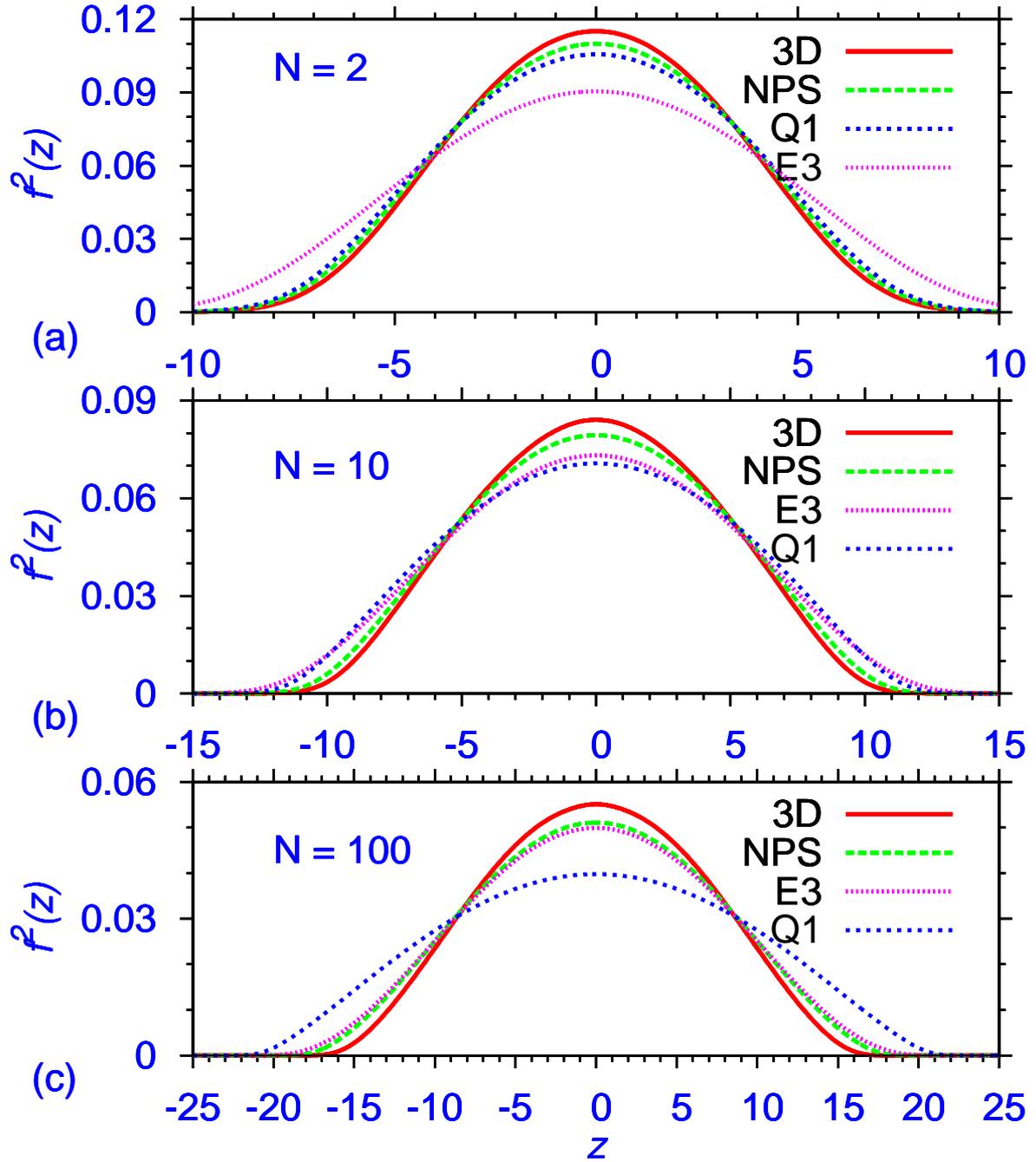}}
 \end{center} \caption{Normalized density $f^2(z)$
($\int_{-\infty}^{\infty}f^2(z)dz=1 $) along the axial $z$ direction for
(a) {2, (b) 10, and (c) 100} Fermi atoms from a 
solution of 3D equation
(\ref{gle2}) denoted 3D, the full one-dimensional equations
(\ref{npse1d1}) and (\ref{al1d1}) denoted NPS, the quasi-1D equation
(\ref{npse1d2}) denoted Q1, and the effectively
 3D 
equation 
(\ref{npse1d3}) denoted  E3 in the unitarity limit using parameters 
$\xi=0.44$, $\lambda_1=0.1,$ and $\lambda_2=1$. } 
\label{fg3}
\end{figure}

For a cigar-shaped superfluid 
we solve four sets of equations in the unitarity limit: (a) the  3D 
equation (\ref{gle2}), 
(b) the complete reduced 1D equations
(\ref{npse1d1}) and (\ref{al1d1}), (c) the approximate quasi-1D
equation (\ref{npse1d2}), and (d) the approximate effectively 3D 
equation (\ref{npse1d3}). We use the parameters $\xi =0.44 $ \cite{mc}, 
and a highly cigar-shaped trap with $\lambda_1=0.1$ and $\lambda_2=1$.
In figure \ref{fg3} we illustrate the results of 
our calculation by plotting the linear density profile $f^2(z)$ vs. $z$ 
of the four sets of calculations for fermion number 
{$N=2,10$ and 100.} 

From figure \ref{fg3} we find that 
the   1D approximate calculations are good approximations to the 
solution of the 3D equation (\ref{gle2}). However, some features of the 
different approximate models are worth commenting. All approximate 1D 
models lead to a density smaller than that obtained 
from the 3D equation 
(\ref{gle2}). The density obtained from  (\ref{npse1d1}) and 
(\ref{al1d1}) provide the best approximation to the exact density for 
all  $N$. For small $N$, the quasi-1D model (\ref{npse1d2})
provides a better  approximation to the exact result than the 
effectively 3D model (\ref{npse1d3}). The opposite happens for large 
values of $N$. For an intermediate value of $N$, the quasi-1D model 
(\ref{npse1d2}) and the
effectively 3D model (\ref{npse1d3}) could produce similar results.  

\subsection{3D-2D crossover}

\begin{figure}[tbp]
\begin{center}
{\includegraphics[width=\linewidth]{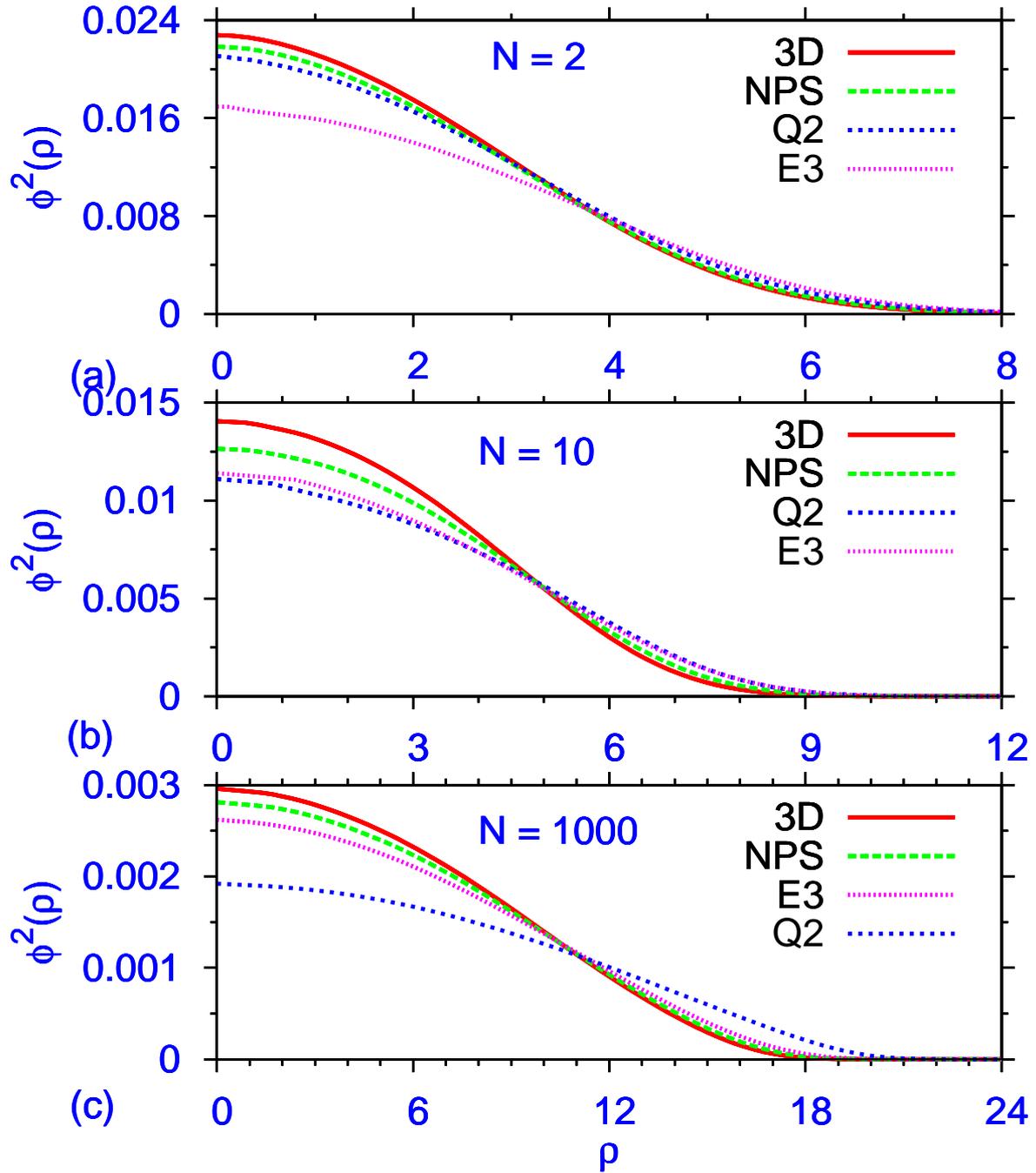}}
 \end{center}
\caption{Normalized density $\phi^2(\rho)$ ($2\pi\int_0^\infty \rho 
\phi^2(\rho)d\rho =1$)
along the radial  $\rho$ direction for (a) 
{2, 
(b) 10, and (c) 1000} Fermi atoms from a solution of 
3D equation 
(\ref{gle2}) denoted 3D, 
the full two-dimensional  equations {(\ref{npse5d}) and 
(\ref{al5d})} 
denoted NPS, 
the quasi-2D equation {(\ref{npse6d})} denoted  Q2, and 
the 
effectively  3D 
equation 
{(\ref{npse7d})} denoted  E3
in the unitarity limit using parameters
$\xi=0.44$, $\lambda_1=1,$ and $\lambda_2=0.1$.} 
\label{fg4}
\end{figure}

For a disk-shaped superfluid we again solve for sets of equations: (a) 
the 3D equation (\ref{gle2}), (b) the complete reduced 2D equations 
(\ref{npse5d}) and  (\ref{al5d}), (c) the quasi-2D equation 
(\ref{npse6d}), and (d) the effectively 3D equation (\ref{npse7d}).
We use the parameters $\xi =0.44$ in the unitarity limit and 
$\lambda_1=1,$ and $\lambda_2=0.1$ for a {disk}-shaped 
trap.
In figure \ref{fg4} we present the results of our calculation by 
plotting 
the radial density profile $\phi^2(\rho)$ vs. $\rho$ of the four sets of 
calculations for fermion numbers {$N=2,10$, and 1000.}

From  figure \ref{fg4} we find that the three 2D approximate models 
could 
be good approximations to the solution of the 3D equation (\ref{gle2}).  
Again, as in the cigar-shaped superfluid, the full 3D model (\ref{gle2})
produces the largest density profile with the complete 2D 
equations (\ref{npse5d}) 
and  (\ref{al5d}) providing the best approximation to it for all values 
of $N$. For small $N$, the quasi-2D model (\ref{npse6d}) produces better 
approximation to the exact result than the effectively 3D model 
(\ref{npse7d}). The opposite happens for large $N$. For an intermediate 
$N$ these two latter approximations could produce similar results. 

An interesting result from our calculations with  cigar and  
disk-shaped superfluid is that,  for {$N>100$}, the 
quasi-1D model 
(\ref{npse1d2}) 
and the quasi-2D model (\ref{npse6d}) are {poorer} 
approximations than the 
effectively 
3D models (\ref{npse1d3}) and (\ref{npse7d}), respectively. For 
experimental purpose {$N=100$} represent  a  small 
number atoms.
Hence for phenomenological applications the
effectively 3D models (\ref{npse1d3}) and (\ref{npse7d}) with nonlinear terms 
with power 9/5 and 2 should 
be used. Note that these powers are different from the power 7/3 in the 
original 3D 
equation (\ref{gle2}). The quasi-1D model (\ref{npse1d2}) and 
quasi-2D model (\ref{npse6d}) for cigar- and disk-shaped superfluids 
with nonlinear terms of power 7/3 effective for a small 
number of fermions are useful for academic interest.

\section{Conclusion}
\label{VII}

We have suggested a time-independent Schr\"odinger equation 
for a Fermi superfluid at unitarity [(\ref{gle2})]
by minimizing  its energy functional. This equation can also 
be derived as an Euler-Lagrange equation of an appropriate Lagrangian
density. In a cigar-shaped superfluid assuming a Gaussian form for the 
order parameter, and integrating over the transverse variables we have 
derived an effective nonlinear nonpolynomial 1D equation for the Fermi 
superfluid at unitarity [(\ref{npse1d1}) and (\ref{al1d1})].  This 
complex equation is simplified in the 
limit of small and large atom numbers when it reduces to a nonlinear 
equation with power-law nonlinearity. The equation for small atom 
number $N$ has the same nonlinear structure as the original 3D equation 
and is called  quasi-1D model [(\ref{npse1d2})], whereas the 
equation for large $N$ has a 
distinct nonlinearity and is called effective-3D model 
[(\ref{npse1d3})]. For 
phenomenological application the  effective-3D model seems 
quite attractive.  

In a disk-shaped superfluid assuming a Gaussian form for the
order parameter, and integrating over the axial variable we also
derived an effective nonlinear nonpolynomial 2D equation for the Fermi
superfluid at unitarity [(\ref{npse5d}) and (\ref{al5d})].  This 
complex equation is simplified in the 
limit of small and large atom numbers when it reduces to a nonlinear
equation with power-law nonlinearity. The equation for small atom
number $N$ has the same nonlinear structure as the original 3D equation
and is called  quasi-2D model [(\ref{npse6d})], whereas the 
equation for large $N$ has a 
distinct nonlinearity and is called effective-3D model 
{[(\ref{npse7d})].} 
The quasi-2D model has the same nonlinearity as the original 3D 
equation, whereas the effective-3D model produces a different 
nonlinearity. Again the effective-3D model is attractive 
for phenomenological application producing very good results. 
All the above models have been studied by a numerical solution of the 
model equations. 

\ack

We thank Prof. Flavio Toigo for useful comments.
S.K.A. was partially supported by FAPESP 
and CNPq (Brazil), and the Institute for Mathematical Sciences of
National University of Singapore.  Research was partially done when 
S.K.A. was visiting the Institute for Mathematical Sciences of
National University of Singapore in 2007.
L.S. has been partially supported by GNFM-INdAM and Fondazione CARIPARO.


\section*{References}

\begin{thebibliography}{99}

\bibitem{CROV} Eagles D M 1969 \PR {\bf 186} 456 

Randeria M, Duan J-M  and Shieh L-Y   1989 \PRL   {\bf 
62} 981 

 Nozieres P and  Schmitt-Rink S 1985 {\it  J. Low Temp. 
Phys.} {\bf 59} 195  

Adhikari S K, Casas M, Puente A, Rigo A, Fortes M, Solis M A, de Llano 
M, 
Valladares A A and  Rojo O  2000 \PR B {\bf 62} 8671

\bibitem{greiner} Greiner M, Regal C A and  Jin D S 2003
{\it Nature}  {\bf 426} 537 

\bibitem{regal}  Regal C A, Greiner M and Jin D S 2004 \PRL
 {\bf 92} 040403 

\bibitem{kinast}  Kinast J,  Hemmer S L,  Gehm M E,
 Turlapov A  and  Thomas J E 2004 \PRL  {\bf 92} 
150402 

\bibitem{zwierlein}  Zwierlein M W {\it et al.} 2004 \PRL
 {\bf 92} 120403 

 Zwierlein M W, Schunck C H,  Stan C A,
 Raupach S M F and  Ketterle W 2005 \PRL
 {\bf 94} 180401 

\bibitem{chin}  Chin  C {\it et al.} 2004 {\it Science} {\bf 305}
1128 

 Bartenstein  M {\it et al.} 2004 \PRL
{\bf 92} 203201 

\bibitem{stringa-fermi} 
 {Giorgini S,  Pitaevskii L P and  Stringari S 2008  {\it 
Rev. 
Mod. Phys.} {\bf 80} 1215 }

\bibitem{uexp} O'Hara K M {\it et al.} 2002 {\it Science} {\bf 298} 
2179 

 Regal C A {\it et al.} 2003 {\it Nature}  {\bf 424} 47 

Strecker K E {\it et al.} 2003 \PRL  {\bf 91} 080406  
 
Jochim S 
{\it et al.} 2003 {\it  Science} {\bf 302} 2101 

  Gehm M E,  Hemmer S L,  Granade S R,  
 O'Hara K M  and 
 Thomas J E \PR  A 2003 {\bf 68} 011401(R) 

\bibitem{uth}Kokkelmans S J J M F,  Milstein J N,  
Chiofalo M L, 
 Walser R  and  Holland M J 2002 \PR  A {\bf 65} 053617 

\bibitem{mc}  Astrakharchik G E,  Boronat J,  Casulleras J  and  
Giorgini S 2004 \PRL  {\bf 93} 200404 

Carlson J,  Chang S-Y,  Pandharipande V R and 
Schmidt K E  2003 \PRL {\bf 91} 050401 

Chang S-Y,  Pandharipande V R,  Carlson J and 
Schmidt K E 2004 \PR   A {\bf 70} 043602 
 
Engelbrecht J R, Randeria M and S\'a de Melo C A R 1997 \PR B {\bf 55} 15153 

Perali A,  Pieri P  and  Strinati G C 2004 \PRL {\bf 93} 100404

\bibitem{baker} Baker G A Jr 1999 \PR C
 {\bf 60} 054311 

Baker G A Jr 2001 {\it Int. J.\ Mod.\ Phys.\ B } {\bf 15} 1314 

 Heiselberg H \PR A 2001 {\bf 63} 043606

\bibitem{fguni}  Bulgac A and  Bertsch G F 2005 \PRL
 {\bf 94} 070401 

 Stringari S 2004 {\it Europhys. 
Lett.}  {\bf 65} 749 

  Bausmerth I,  Recati A and  Stringari S 2008 \PRL
 {\bf 100} 070401 

   Hu H,  Liu X-J 
and 
Drummond P D 2007 \PRL
 {\bf 98} 060406 

\bibitem{ginzburg}  Ginzburg V L and  Landau L D 1950
{\it Zh. Eksp. Teor. Fiz.}  {\bf 20} 1064 

\bibitem{landau2}  Landau L D and  Lifshitz E M (1987)
{\it Statistical Physics, Part 2: Theory of the Condensed State,
Course of Theoretical Physics, vol. 9}
(London, Pergamon Press), Ch. 5.

\bibitem{leggett}  Leggett A J 2006 {\it Quantum Liquids}
(Oxford, Oxford Univ. Press), Ch. 5.

\bibitem{kim-zubarev}  Kim Y E and  Zubarev A L 2004 \PR A 
 {\bf 70} 033612 

 Kim Y E and  Zubarev A L 2005 \PR A
 {\bf 72} 011603(R) 

 Kim Y E and  Zubarev A L 2004 \PL A 
 {\bf 397} 327

 Kim Y E and  Zubarev A L 2005 \jpb
  {\bf 38} L243

\bibitem{manini05} 
 Manini N  and Salasnich L 2005 \PR
 A {\bf 71} 033625  

 Diana G,  Manini N and 
 Salasnich L 2006  \PR  A {\bf 73} 065601 

\bibitem{sala-josephson}  Salasnich L,  Manini N and  Toigo F 2008 
\PR A {\bf 77} 043609 

\bibitem{ska1} Adhikari S K 2008 \PR  A {\bf 77} 045602 

\bibitem{HY}  Lenz W  1929 {\it Z. Phys.} {\bf 56} 778 

 Huang K and Yang C N  1957 \PR  {\bf 105} 767 

 Lee T D  and  Yang C N 1957 \PR    {\bf 105} 1119 

\bibitem{sala-new}  Salasnich L 2008 e-preprint arXiv:0804.1277  
to be published in Laser Phys.

\bibitem{rupak}  Rupak G and  Sch\"afer T 2008 
e-preprint arXiv:0804.26782v2. 

\bibitem{son}  Son  D T and  Wingate M 2006 {\it
Ann. Phys. (N.Y.)}  {\bf 321} 197 

\bibitem{sadhan} Adhikari S K and Salasnich L 2008 \PR A {\bf 78} 043616 

\bibitem{FN} Blume D,  von Stecher J  and  Greene C H 2007 \PRL
{\bf 99} 233201 

von Stecher J,  Greene C H and  Blume D 2008 \PR   A {\bf 77} 043619 

\bibitem{FN1}  Chang S Y and  Bertsch G F 2007 \PR   A {\bf 76} 021603(R) 

\bibitem{sala-odlro} Salasnich L,  Manini N and  Parola A 2005
\PR A {\bf 72} 023621

Salasnich L 2007 \PR A {\bf 76} 015601  

\bibitem{giorgini-odlro}  Astrakharchik G E,  Boronat J,  Casulleras J
and  Giorgini S 2005 \PRL
{\bf 95} 230405 

\bibitem{kohn1} Hohenberg  P and  Kohn W  (1964)  \PR {\bf 136} B864 

Kohn W  1999 {\it Rev. Mod. Phys.} {\bf 71} 1253 

Dreizler R M and 
and  Gross E K U 1990  {\it Density Functional Theory; An Approach to 
the Quantum Many-Body Problem} (Berlin, Springer)

\bibitem{kohn2} Oliveira L N,  Gross E K U and  Kohn W 1988 \PRL
 {\bf 60} 2430 

\bibitem{kohn} Kohn  W and  Sham L J 1965  \PR {\bf 140} A1133 

\bibitem{von}  von Weizs\"acker C F 1935 {\it 
Z. Phys.} {\bf 96} 431 

\bibitem{tosi} March  N H and  Tosi M P 1973 
{\it Ann. Phys. (NY)} {\bf 81} 414 
 
 Vignolo P,  Minguzzi A and  Tosi M P 2000 \PRL  {\bf 85}  2850 

\bibitem{SKA1} 
 Adhikari S K 2004 \PR A 
{\bf{70}} 043617   


\bibitem{zaremba} Zaremba E and  Tso H C 1994 \PR 
 B {\bf 49} 8147 

\bibitem{sala-gradient}  Salasnich L 2007 \JPA 
 {\bf 40} 9987 

\bibitem{strinati}  Pieri P and  Strinati G C 2003 \PRL 
 {\bf 91} 030401

\bibitem{dicastro}  De Palo S,  Castellani C,  Di Castro C and
 Chakraverty B K 1999 \PR   B {\bf 60} 564 

\bibitem{cowell}  Cowell S,  Heiselberg H, Mazets I E, 
Morales J,  Pandharipande V R and  Pethick C J 2002 \PRL
{\bf 88} 210403 

\bibitem{hei}  Heiselberg H 2004 \jpb  {\bf 37} S141  

\bibitem{sala-fermi}  Salasnich L 2000 {\it  J. Math. Phys.} {\bf 41} 
8016 

\bibitem{bulgac}  Bulgac A 2007   \PR A {\bf 76} 040502(R) 

\bibitem{toigo} {Salasnich L and Toigo F 2008 \PR A {\bf 
78} 053626} 


\bibitem{blume} von Stecher J,  Greene C H and  Blume D 2007 \PR   A 
{\bf 76} 063613




\bibitem{sala-npse}  Salasnich L,  Parola A  and  Reatto L 2002 
\PR A 
 \textbf{65} 043614  

\bibitem{sala-tonks(a)}  Salasnich L,  Parola A  and  Reatto L 2004  \PR 
A \textbf{70} 013606 

\bibitem{sala-tonks(b)} Salasnich L,  Parola A  and  
Reatto L 2005
\PR A 
 \textbf{72} 025602 

\bibitem{sala-npse-2cp}  Salasnich L 
and  Malomed B A 2006 \PR A   \textbf{74}
053610 

\bibitem{sala-npse-tr}  Salasnich L,  Cetoli A,  Malomed B A, 
 Toigo F  and  Reatto L 2007  \PR A \textbf{76} 013623 

\bibitem{sala-npse-vor}  Salasnich L,  Malomed B A  and  Toigo F 2007 
\PR A {\bf 76} 063614 

\bibitem{adhi}  Adhikari S K 2005 \PR A
{\bf 72} 053608 

 Adhikari S K and  Malomed B A 2007 \PR A
 {\bf 76} 043626 

 Adhikari S K and  Malomed B A 2006 \PR A
{\bf 74} 053620

 Adhikari S K 2006 \PR A
{\bf 73} 043619



\bibitem{adhi2}  Adhikari S K 2007 \JPA 
 {\bf 40} 2673 

Adhikari S K 2006 {\it Eur. 
Phys. J. D}  {\bf 40} 157 

Adhikari S K 2006 {\it Laser Phys. Lett.} {\bf 3}   605 

Adhikari S K  1979 \PR C {\bf 19} 1729

\bibitem{coz}  Cozzini M and  Stringari S 2003 \PRL
{\bf 91} 070401 

\bibitem{new} Adhikari S K and  Salasnich L 2007 
 \PR A  {\bf 75} 053603 

\bibitem{num}  Koonin S E and  Meredith D C 1990
 {\it Computational Physics 
Fortran Version}, (Addison-Wesley, Reading) 

\bibitem{sala-numerics}  Cerboneschi E,  Mannella R,  Arimondo E
 and Salasnich L 1998 {\it Phys. Lett.} A {\bf 249} 495  

Salasnich L,  Parola A
and  Reatto L 2001 {\it  Phys. Rev. A} {\bf 64} 023601   

\bibitem{cn}  Adhikari S K and  Muruganandam P 
2002 \jpb  {\bf 35} 2831 




\end{thebibliography}
\end{document}